\documentclass[aps,prl,twocolumn,longbibliography,a4paper]{revtex4-1}

\usepackage{graphicx}
\usepackage{braket}
\usepackage{amsmath}
\usepackage{amssymb}
\usepackage{amsthm}
\usepackage{MnSymbol}

\DeclareMathOperator{\Tr}{Tr}

\begin{document}


\title{Spatial Quantum State Tomography with A Deformable Mirror}
\author{K. S. Kravtsov$^{1,2,3}$}\email{kk@quantum.msu.ru}
\author{A. K. Zhutov$^2$}
\author{S. P. Kulik$^2$}
\affiliation{
	$^1$ Prokhorov General Physics Institute of the Russian Academy of Sciences, 119991 Moscow, Russia \\
	$^2$ MSU Quantum Technology Centre, 119991 Moscow, Russia\\
	$^3$ National Research University Higher School of Economics, 101000 Moscow, Russia\\
}

\date{\today}
\begin{abstract}
Quantum tomography is an essential experimental tool for testing any quantum technology implementation.
Transverse spatial quantum states of light play a key role in many experiments in the field of quantum information as well as in free-space optical communications. In this paper we propose and experimentally demonstrate the tomography of spatial quantum states with a deformable mirror. It may be used to significantly outperform the conventional method with a spatial light modulator (SLM) in terms of speed and efficiency, at the same time providing polarization insensitive operation.
\end{abstract}

\maketitle\nopagebreak
\section{Introduction}
The tomography of a quantum state is a standard procedure of determining an unknown quantum state by performing a series of measurements over a large number of its copies~\cite{WF89,JKM01,REK04}. This procedure is widely studied both theoretically and experimentally.
In this paper we focus on the tomography of transverse spatial quantum states of light, which are often used as a model qudit system~\cite{MVW01,LPB02,MTT07}.

Unlike the two-dimensional case, where efficient experimental tools for quantum measurements are readily available, the realization of general measurements in higher dimensions is typically a much more complex problem. In some cases, e.g. for a composite qubit system, it is still easy to perform a {\it factorized} measurement~\cite{SKS18}, effectively reducing the overall task to single qubit measurements. On the contrary, higher dimensional {\it spatial} quantum states of light use degrees of freedom of a single particle, and so there is no universal and efficient solution for their measurement.

A lot of efforts were targeted at the development of so-called mode sorters~\cite{BLC10,HML15,RGM18,FRC19}. They may be used for separation of different spatial modes that represent a natural ({\it computational}) basis for the quantum system in question. However, mode sorters formally enable only a single type of measurement, namely, measurements in the computational basis, that is obviously insufficient for a full state tomography. A typical solution for a general projective measurement is the use of a liquid crystal-based SLM for mode transformation followed by a single-mode fiber filter that performs projection of the resulting field onto the fundamental mode~\cite{MVR04,BQT15}. Although this method has been the standard technique of projective measurements in the spatial mode space, the SLM-based measurement technique has a number of significant drawbacks, namely associated with its low switching speed, poor efficiency, and polarization sensitivity.

The goal of this paper is to show that a measurement SLM may be effectively replaced by a MEMS deformable mirror, which can be switched at a much higher rate and at the same time provide nearly lossless mode transformation and polarization insensitive operation.
In the result, quantum tomography of substantially bright sources can be potentially performed at millisecond time frames and even faster, which is far from being tractable with conventional SLMs.
This may find applications in real-time tomography of turbulent atmospheric channels and other non-stationary experimental environments.

We experimentally demonstrate a deformable mirror-based tomography in a 4-dimensional Hilbert space, that yields adequate fidelity of the reconstructed quantum states. 

\section{Quantum tomography}

The goal of quantum tomography is to reconstruct an unknown density matrix $\rho$ by a series of measurements performed on copies of the quantum state in question.
According to the Born's rule the probability of observing an outcome $\gamma$ for a positive operator-valued measure (POVM) $M$ with elements $\{M_\gamma\}$ is $P(\gamma\,|\rho) = \Tr (M_\gamma \rho)$. Experimentally obtained probabilities for a sufficient number of different POVM elements allow to calculate the desired $\rho$.  
As mentioned before, the realistic measurement technique is limited to projective measurements only, so a general POVM formalism may be reduced to the following form of $M_\gamma$:  $M_\gamma = \ket{P_\gamma}\bra{P_\gamma}$.

There has been a lot of progress in quantum state reconstruction methods recently, including establishing of adaptive~\cite{S16}, self-guided~\cite{F14}, compressed-sensing and neural-network enhanced~\cite{ATJ19,PKB20} and many other techniques including the so-called shadow tomography~\cite{HKP20}. In  this work we primarily focus not on the tomography protocol itself but rather on the experimental tool that is used for implementation of the protocol. Therefore, for this proof of concept demonstration we use the most straightforward protocol, based on the maximum likelihood estimation. In our understanding the proposed experimental tool is efficient with tomography protocols, where only a fixed number of measurements is required. That basically rules out all truly adaptive approaches, where a continuous shift of the measurement projection is implied. However, some strategies such as neural network-based correction for state preparation and measurement errors~\cite{PKB20}, will definitely improve the obtained results. To keep the paper succinct and the results clear, in the further text we do not make any additional assumptions about the protocol used, and show an example of the whole scheme implementation with the most basic tomography protocol, described in detail later. More advanced versions of quantum state tomography may be built on top of the presented demonstration.

The minimum number of projectors required for a complete tomography in a $d$-dimensional Hilbert space is $d^2$, which may be chosen as the set of the symmetric, informationally complete (SIC)-POVM elements. That would give the most uniform distribution of the expected reconstruction fidelity among all possible quantum states. In this paper we use a redundant set of $d\,(d+1)$ projectors onto elements of mutually unbiased bases (MUBs), which preserves the same coverage uniformity as SIC-POVM, but results in a much more symmetric and easy-to-implement array of the deformable mirror states.

When it gets to experimental realization of projective measurements especially with a continuous deformable mirror, actual projectors may substantially differ from the designed ones in terms of both efficiency and fidelity. In fact, lossless projections onto MUBs elements is impossible to realize even with a perfect phase mask.  Thus, before the actual state tomography one needs to perform the so called {\it detector tomography}, to find the actual projectors that take place in the experiment~\cite{LFC09,BKM15}.

\section{Experimental realization}

In our demonstration we work with the following set of Hermite-Gaussian modes constituting the computational basis: HG$_{00}$, HG$_{01}$, HG$_{10}$, and HG$_{11}$, so the dimension of the associated Hilbert space is $d=4$.
Such a choice is made due to the rectangular geometry of the MEMS mirror actuator array, which well matches that of the HG modes. Computational bases with other symmetry types, e.g. a circular one for Laguerre-Gaussian or orbital angular momentum (OAM) modes, are in general less effective here as they would require more independent pixels to approximate corresponding MUBs elements with a rectangular grid. After all, the obtained tomography results may be easily converted to any other basis representation as all of them are completely equivalent~\cite{DA92}. As a figure of merit for state reconstruction we use the {\em fidelity}, defined as
\begin{equation}
F = \left[\mathrm{Tr\left(\sqrt{\sqrt{\rho}\rho'\sqrt{\rho}}\right)}\right]^2.
\end{equation}

\begin{figure}
	\includegraphics[width=\columnwidth]{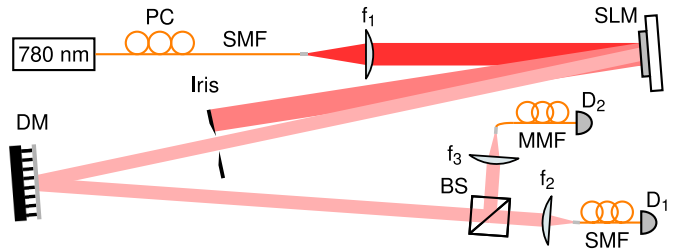}
	\caption{Experimental setup. PC --- polarization controller, SMF --- single-mode fiber, DM --- deformable mirror, BS --- symmetric beam splitter, MMF --- multi-mode fiber; D$_{1,2}$ --- single photon avalanche detectors. }\label{fig_DM_exp}
\end{figure}

The experimental setup used in our realization is shown in Fig.~\ref{fig_DM_exp}. The state preparation is performed with an SLM using the conventional mode synthesis technique described in~\cite{BBS13}. The zero order diffraction is blocked by an iris diaphragm placed at a distance of around 1~meter from the SLM. The tomography stage is implemented with a Boston Micromachines Mini-3.5 deformable mirror (DM) consisting of 32 active elements arranged as a 6x6 matrix with missing corners. The resulting light field is focused into the single-mode fiber (SMF), delivering it to the single photon avalanche detector (SPAD). As the rate of photons in the synthesized modes varies with the mode, a reference channel with a multi-mode fiber (MMF) is added by placing a symmetric beamsplitter in the optical path.

The working wavelength is 780~nm, and the beam waist parameter of HG beams is $w_0 = 0.9$~mm. As the distance between the SLM and the DM of 1.92~m is comparable with the Rayleigh range of $z_\mathrm{R} = 3.26$~m, the SLM hologram was adjusted to set the beam waist right at the DM plane. Besides the apparent changes of the phase curvature and the hologram size, this also requires proper accounting of the Gouy phase to get correct relative phases for superpositions of modes. Thus, with the help of the SLM we can synthesize an arbitrary spatial quantum state at the DM plane. It is considered to be the input of the further {\em tomography} step. The synthesis technique is assumed to be ideal, as it gives the fidelity of at least 0.99.

\begin{figure}
	\includegraphics[width=\columnwidth]{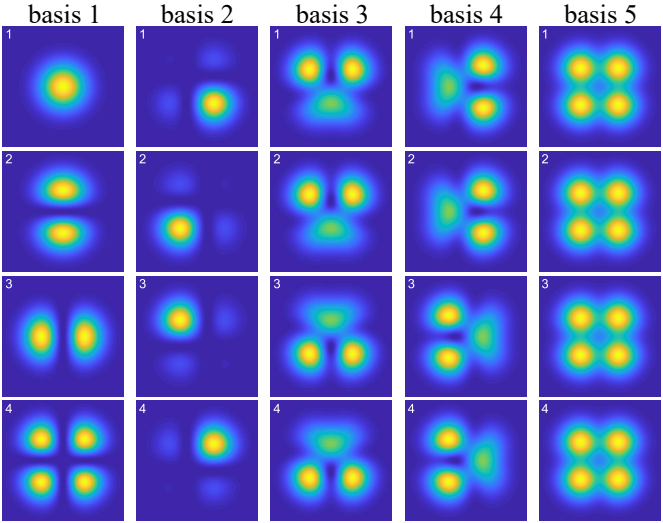}
	\caption{Spatial distribution of amplitudes for all 20 MUB elements used for the tomography: 5 MUBs with 4 elements in each. Identically looking patterns differ by the phase distribution, which is not shown here.}\label{fig_DM_MUB_elements}
\end{figure}

The tomography model that we exploit here would require measuring projections of the incoming states onto 20 MUBs elements shown in Fig.~\ref{fig_DM_MUB_elements}. So the ideal tomography device would losslessly convert any of the MUBs elements into the fundamental mode perfectly matching that of the SMF. In other words, if we launch the signal backwards, the tomography device is required to be capable of converting the SMF output into any of the MUBs elements. This can be easily performed with an SLM, as it is typically done in many experiments. However, it has the already mentioned drawbacks of significant signal loss, quite limiting switching rates, and strong polarization sensitivity.

To overcome these limitations we use a DM instead. The DM has its own flaws, and the most obvious is its poor spatial resolution. Indeed, it is impossible to synthesize an arbitrary spatial state from the SMF output using a DM. However, the good news is that to perform the tomography we do not need to project exactly on MUBs elements. As long as we know what we project upon, we will be able to reconstruct the states with the high fidelity even if the projectors are quite far from the desired MUBs elements. To learn the actual projectors we perform the detector tomography. It is completely equivalent to tomography of all quantum states that we would get after the DM if we launch signal backwards, i.e. from the SMF onto the DM.

What we do experimentally, is we synthesize all 20 MUBs elements one by one, and optimize the shape of the deformable mirror to achieve the best coupling of these 20 MUBs elements into the SMF. The basic idea behind this is to rectify the phase, thus attaining the best overlap of the resulting field with the SMF mode. Thus, we know the required geometry of the DM states and all optimization that we performed was the adjustment of the amplitudes of the DM shifts. In the end, we get projections upon something roughly resembling the desired MUBs elements. To use them as tomographic measurements we need to know exactly what they are, i.e. we perform the detector tomography.


In general, mode transformation in the deformable mirror may be described by a unitary matrix $M'$ in a larger Hilbert space that includes higher order spatial
modes:
\begin{equation}
	\begin{bmatrix}
		a_1\\ a_2\\ a_3\\ a_4\\ a_5\\ \vdots \\a_n
	\end{bmatrix}
	=
	M'
	\begin{bmatrix}
		b_1\\ b_2 \\ b_3 \\ b_4\\ b_5 = 0\\ \vdots \\ b_n = 0
	\end{bmatrix},
\end{equation}
where we assume that the input state is limited to the dimension of $d = 4$.

Thus, the tomography of DM states accesses only the left upper corner of the whole matrix $M'$:
\begin{equation}
	M'=
	\begin{bmatrix}
		m_{11} & m_{12} & m_{13} & m_{14} & \hdots &\\
		m_{21} & m_{22} & m_{23} & m_{24} & \hdotsfor{2}\\
		m_{31} & m_{32} & m_{33} & m_{34} & \hdotsfor{2}\\
		m_{41} & m_{42} & m_{43} & m_{44} & \hdotsfor{2}\\
		\hdotsfor{6}\\
		\hdotsfor{6}\\
	\end{bmatrix}.
\end{equation}
We will call it $M$. This is a matrix with all eigenvalues $|\lambda_k| \le 1$, because of a possible power leak into higher order modes.

In the experiment we directly (up to the constant splitting ratio and the efficiencies of the detectors) measure probabilities $P_{ij}$ of coupling into the single-mode fiber, where $i$ is the number of the input state
$\ket{\Phi_i}$ and $j$ is the number of the mirror setting; $i,j \in \{1,2\dots 20\}$.
It equals 
\begin{equation}
	P_{ij} = \left|\braket{\Psi_{00}|M_j| \Phi_i}\right|^2,
	\label{eq_dm_prob}
\end{equation}
where $\ket{\Psi_{00}}$ is the fundamental mode of the SMF
and $\ket{\Phi_i}$ --- the MUBs elements.

\begin{figure*}
	\includegraphics[width=\textwidth]{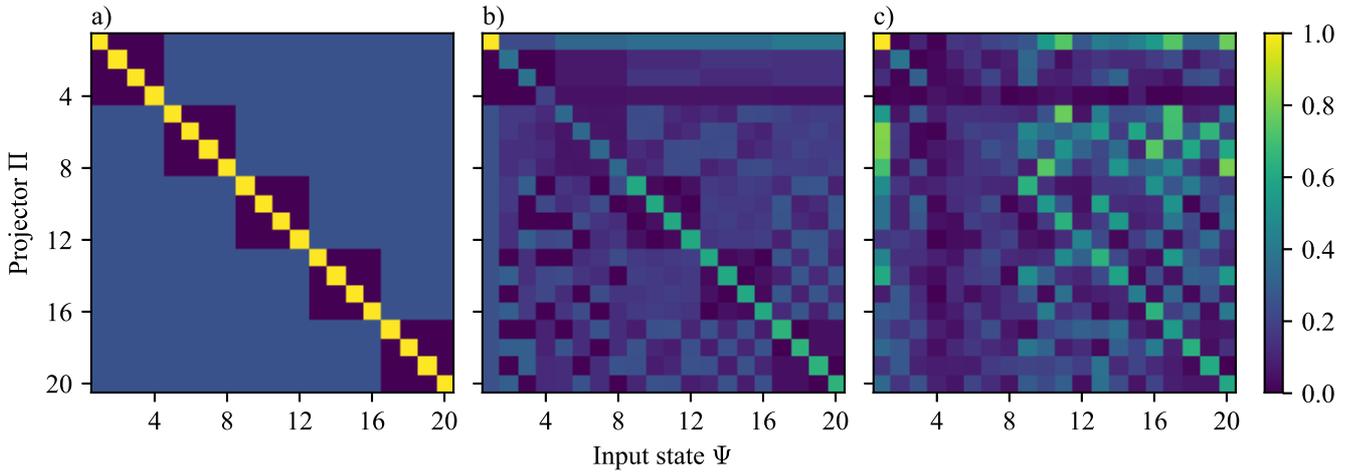}
	\caption{A matrix of measurement results for the MUB-based protocol in a 4-dimensional Hilbert space: the probability of a state $\Phi_i$ to pass through the projector $\Pi_j$ a) ideal projectors onto MUB states $\Phi_j$; b) ideal deformable mirror with the infinite size and resolution; c) actual measurements with our experimental setup.}
	\label{fig_DM_P_ji}
\end{figure*}

If the mirror was the ideal tomography device, it would work such as $M_j^\dagger \ket{\Psi_{00}} =
\ket{\Phi_j}$, so the $j$-th projector would be simply the projector onto $\ket{\Phi_j}$. This would result in a matrix of $P_{ij}$ shown in Fig.~\ref{fig_DM_P_ji}a. In reality, a mirror even with the infinite size and resolution is not capable of such a transformation. Even the perfectly rectified phase of the field does not allow to match the amplitude profiles that leads to significant deviations of probabilities. This scenario of a perfect mirror is simulated in Fig.~\ref{fig_DM_P_ji}b. As one can see, some MUB elements, especially 4 and 5-8 (see Fig.\ref{fig_DM_MUB_elements}) have a smaller intensity overlap with the fundamental mode and, thus, smaller probabilities of passing through. 

The actual results of the detector tomography, i.e. the experimentally measured matrix of $P_{ij}$ is shown in Fig.~\ref{fig_DM_P_ji}c. Despite a significant deviation from the ideal DM, which is expected because we use only a 6x6 pixels DM with a continuous reflective membrane, the general pattern still holds and the projectors are assumed to retain the high order of symmetry to effectively perform the quantum state tomography. The lack of a high spatial resolution together with minor alignment errors results in noticeable deviations from Fig.~\ref{fig_DM_P_ji}b. For example, some non-diagonal elements on the plot are larger than the diagonal ones. The advantage of the demonstrated approach is that this disturbance can not significantly affect the tomography results, as we measure the projectors in the detector tomography step and then use this knowledge for state reconstruction.

To solve the overdetermined system of equations~(\ref{eq_dm_prob}) for $M_j$, it is easier to use a more general formalism.
Let the input state is described by a density matrix $\rho$, which is measured with the particular mirror
setting $j$. Then the measured probability equals
\begin{equation}
	P_{j} = \Tr \left( \rho \cdot M_j^\dagger \ket{\Psi_{00}}\bra{\Psi_{00}} M_j\right).
\end{equation}
This equation, being effectively a Hilbert-Schmidt inner product for the two matrices, may be rewritten using the vectorized notation~\cite{GTW09,W15}
\begin{equation}
	P_{j} = \Big\llangle \rho \; \Big| M_j^\dagger \ket{\Psi_{00}}\bra{\Psi_{00}} M_j \Big\rrangle.
	\label{eq_dm_prob2}
\end{equation}

In the detector tomography step the input states are known and are described by the $\rho_i = \ket{\Phi_i}\bra{\Phi_i}$, while the
measurement matrix $\Pi_j = M_j^\dagger \ket{\Psi_{00}}\bra{\Psi_{00}} M_j$ is to be found. By the construction, this
matrix $\Pi_j$ is
an improperly normalized projector, i.e. has
only one non-zero eigenvalue corresponding to the eigenvector $\ket{P_j} = M_j^\dagger \ket{\Psi_{00}}$. The
eigenvalue itself shows the efficiency of the projection. For each $j$ one needs to solve the overdetermined system of linear equations with varying $i$:
\begin{equation}
	P_{ij} = \Big\llangle \rho_i \; \Big| M_j^\dagger \ket{\Psi_{00}}\bra{\Psi_{00}} M_j \Big\rrangle.
\end{equation}
The system is solved in the sense of a least-squares solution. Due to the presence of experimental errors especially in the case of the overdetermined system the resulting matrix $\Pi_j$ contains more than one non-zero eigenvalue. In the detector tomography step the minor eigenvalues are dropped to explicitly keep the form of this operator as $\Pi_j = \ket{P_j}\bra{P_j}$.

The quantum state tomography is the reverse of the detector tomography: knowing $\Pi_j$ and $P_j$ for all $j$s, one needs to solve the system~(\ref{eq_dm_prob2}) for the unknown $\rho$.

It is quite apparent that almost any $d^2 = 16$ different DM states are sufficient for the qudit tomography. However, some sets are more optimal than others. The quantitative measure may be the ratio between the largest and the smallest singular values of the matrix formed by the set of vectorized measurement matrices $\big| \Pi_j \big\rrangle$: $\eta = \max \lambda_k /\min \lambda_k$. Indeed, if $\eta$ is close to 1, inverting the system~(\ref{eq_dm_prob2}) appears to yield the smallest uncertainty due to the experimental noise. On the contrary, very large $\eta$s lead to the strong propagation of small experimental noise into the solution. 

The ideal case of $\Pi_j = \ket{\Phi_j}\bra{\Phi_j}$ corresponds to $\eta = \sqrt{5} \approx 2.2$. The infinite resolution mirror yields $\eta \approx 5.0$; an ideal 6x6 mirror with independent pixels and the same pixel size as in the experiment shows $\eta \approx 10.5$. The measured matrix yields $\eta \approx 33$, which is not so far from the idealized case.  The main difference between the latter two is due to the large pixel-to-pixel cross-talk in our DM. This is an
unavoidable drawback of the continuous mirror-based DMs, where adjacent pixels are significantly coupled through the reflective membrane. Although, the corresponding $\eta$ could be potentially calculated via thorough modeling of the mirror, it is not realistic because we do not know much about its inner structure. We can only conclude that the aggregate effect of this cross-talk and all other experimental imperfections, including ever-present minor alignment errors, leads to the 3-fold $\eta$ increase, which looks rather reasonable.
Finally, as mentioned before, the value of $\eta$ has no direct effect on the obtained fidelities. A noiseless and perfectly repeatable experiment would always yield perfect results anyway. A large $\eta$ only means stronger noise propagation into the tomographic results.

After the DM calibration matrix (Fig.~\ref{fig_DM_P_ji}c) has been exprerimentally measured, we calculate all 20 measurement matrices $\Pi_j$, i.e. obtain the result of the detector tomography. Then we use these data for quantum state tomography to reconstruct the supposedly unknown input states $\rho$ after measuring them with all 20 projectors $\Pi_j$ each. In the experiment, however, we know exactly what these states are because we synthesize them with the SLM. Therefore we can calculate the fidelity between the reconstructed and the actual states.

\begin{figure}
	\includegraphics[width=\columnwidth]{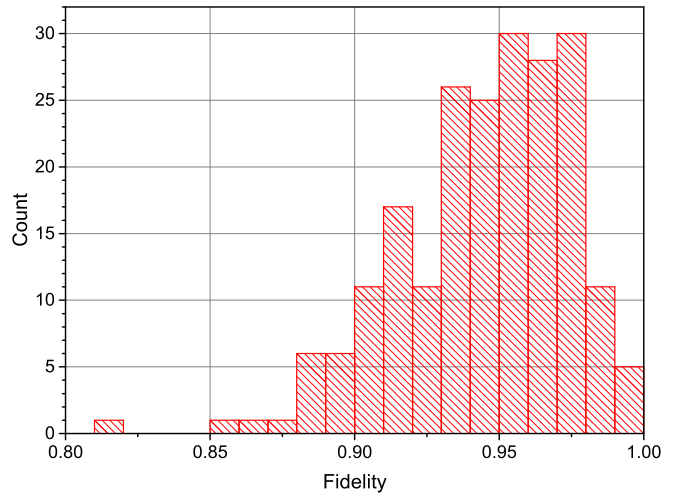}
	\caption{Distribution of the reconstructed states fidelity for a set of 210 random pure quantum states.}
	\label{fig_DM_fid_histo}
\end{figure}

First of all we accomplished the tomography of the 20 $\ket{\Phi_i}$  states that yields the average fidelity with the actual states of 0.977 and the worst case fidelity of 0.940. Then, random pure qudit states were generated and their density matrices were reconstructed using the tomographic procedure. Figure~\ref{fig_DM_fid_histo} shows the histogram of the fidelity for the measured 210 random quantum states. The obtained fidelities are not particularly high and can not be considered to be the best among similar experiments, however, they are quite typical for spatial state tomography of a dimension of 4. It is a well-known experimental fact that the larger the reconstructed state dimension the smaller are the typical fidelities~\cite{BQT15}. This is mainly due to the experimental errors that propagate into the obtained reconstruction results, as spatial quantum states are quite fragile in this sense. At the same time, it is somewhat offset by the fact that the expected fidelity between two random pure states is $\langle F_{\mathrm rnd}\rangle=1/d$ with the exact probability distribution of $P(F) = (d-1)(1-F)^{d-2}$ \cite{ZS05}, so, for example, the measure of the states having the fidelity with the given one of $F>0.9$ in  a 4-dimensional space is only $10^{-3}$.

\section{Discussion}
The performed experimental demonstration shows the plausibility of the DM-based approach to spatial qudit tomography. The experimentally obtained fidelity of the reconstruction is quite similar to other tomographic experiments with spatial quantum states~\cite{BQT15,PKB20}.

There are three main advantages of using a DM instead of an SLM: 1. efficiency of mode transformation; 2. polarization insensitivity; 3. speed of operation. The gain in efficiency is due to the way a liquid crystal SLM operates. Its reliable use as a phase screen is typically performed in the first order of diffraction, while all other orders, including the 0-th, lead to the signal loss. The phase shifts strongly depend on the polarization, limiting SLMs to polarized inputs only. The DM, on the contrary, reflects nearly 100\% of light, giving in practice virtually no loss, but the pure polarization insensitive phase shift.

The speed of operation of a liquid crystal-based SLM is limited by a few hundred Hz, as these large molecules are not agile enough to move faster under the applied electric field. The MEMS-based devices easily show kHz switching rates and much more~\cite{AGG14,ZFD16} so a properly driven DM may be able to switch states orders of magnitude faster than conventional SLMs. Another good reason for using the DM is a much more reasonable amount of data required to define a new state. In our setup we only need 32 bytes of data to completely describe the states of all 32 available actuators. More recent DMs typically have at least 100  and into a few thousand actuators, which is enough for performing tomography in a much larger Hilbert space than in our demonstration. On the other hand, a typical SLM is a megapixel class device, requiring at least 1~MB of data to define all its pixels. The data should be calculated and then transferred to the device. This by itself takes at least 3 orders of magnitude more time than transferring sub kB data for the DM, and is typically limited by the supported frame rate of the HDMI/DVI interface of a few hundred Hz.

Thus, the DM-based tomography may play a key role for the tomography of non-stationary states, where the high rate of measurements is crucial. For example, it may be used in free-space communication channels to measure the disturbance of the transmitted quantum states as well as in other dynamic experiments, where quantum states vary in time. In the same way, it could be used for the measurement of modal composition of classical bright beams of light, as in classical free-space communications.

While the advantage in the efficiency is not that striking, DMs still may replace SLM's to improve the overall detection efficiency, critical in many quantum technology demonstrations. This especially applies to the tomography of unpolarized sources.

In conclusion, we demonstrated the use of a deformable mirror for transverse spatial state tomography. Being one of the most commonly used qudit systems, spatial quantum states of light play a key role in many experiments in the field of quantum information. Their fast and reliable tomography is the cornerstone of further advancement into the field. The proposed approach allows to perform tomography orders of magnitude faster and yielding higher efficiency than with the conventional SLM-based approach. The experimentally performed tomography of randomly sampled qudit states demonstrated reasonable fidelity of reconstruction.

\begin{acknowledgments}
The development of the main concept and its partial implementation was supported by the RFBR grant No. 17-02-00966. The experiment directly relies on the detection scheme and optimization techniques developed as a part of the implementation of the state support Program of NTI Centers, namely under the  the free-space quantum communications initiative.
\end{acknowledgments}

\end{document}